\documentstyle[graphicx,longtable]{aipproc}
 
\begin{document}
\title{Aspherical Supernova  Explosions: Hydrodynamics, Radiation Transport
 \& Observational Consequences}
\author{P. H\"oflich$^1$, A. Khokhlov $^2$, L. Wang$^3$}
\address{
$^1$ Department of Astronomy, University of Texas, Austin, TX 78681, USA\\
$^2$ Naval Research Lab, Washington DC, USA\\
$^3$ Lawrence Berkeley Lab, 1 Cyclotron Rd, Berkeley, CA 94720, USA
}

\maketitle

\begin{abstract} 
 Core collapse supernovae (SN) are the final stages of stellar evolution in 
massive stars during which the central region collapses, forms a neutron star 
(NS), and the outer layers are ejected. Recent explosion scenarios assumed that 
the ejection is due to energy deposition by neutrinos into the envelope but 
detailed models do not produce powerful explosions. There is new and mounting evidence
for an asphericity and, in particular, for axial symmetry in several supernovae which
may be hard to reconcile within the spherical picture.
  This evidence includes the observed high polarization and its variation with time,
pulsar kicks, high velocity iron-group and intermediate-mass elements material 
observed in remnants,  direct observations of the debris of SN1987A etc.
 Some of the new evidence is discussed in more detail.
 To be in agreement with the observations, any successful mechanism must invoke
some sort of axial symmetry for the explosion.

 As a limiting case, we consider jet-induced/dominated explosions of "classical" 
core collapse supernovae. The discovery of magnetars revived the idea that
a MHD-jet with appropriate properties may be formed at the NS.
 Our study is  based on detailed 3-D hydrodynamical 
and radiation transport models.  We demonstrate the influence of the jet properties and 
of the underlying progenitor structure on the final density and chemical structure.
 Our calculations show that low velocity, massive jets can explain the observations.
 Both asymmetric ionization and density/chemical distributions have been identified
as crucial for the formation of asymmetric photospheres.  Even within the picture
of jet-induced explosion, the latter effect alone
fails to explain early polarization in core collapse supernovae with a massive, hydrogen-rich envelopes
 such as SN1999em. Finally, we discuss observational consequences and tests.
\end{abstract}

\section{Observational Evidence for Asymmetry}

\begin{figure}[ht]
\includegraphics[width=6.8cm,angle=270,clip=]{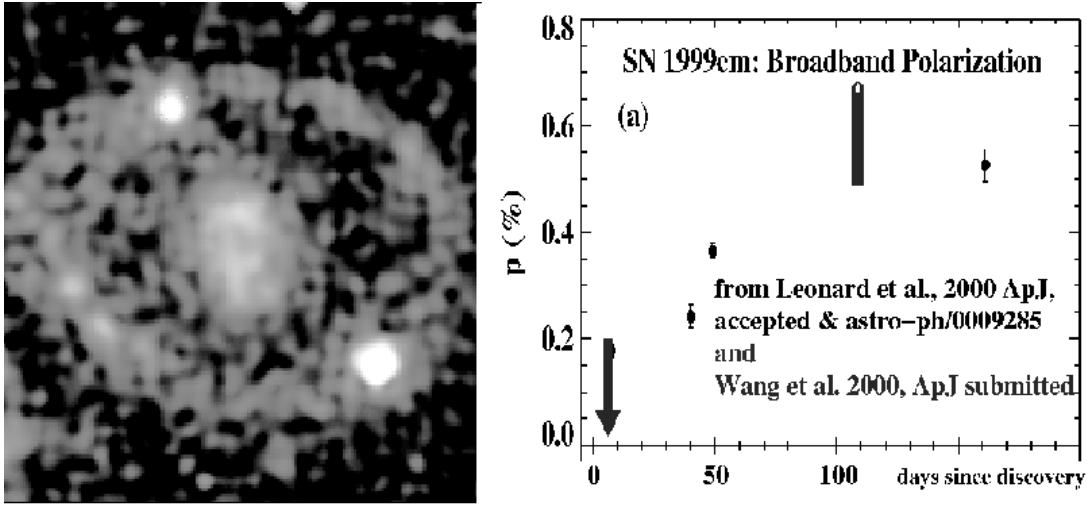}
\caption {
 Observational evidence for asphericity in core collapse supernovae.
The HST image of SN~1987A (left panel) shows the inner debris of the SN-ejecta with an axis ratio of $\approx 2$ and the ring
 on June 11th, 2000 (from Wang et al. 2001). Note that the inner ring
has been formed during the stellar evolution  about 10,000 years before the explosion.
The right panel shows the evolution with time of the linear polarization $P$ in the plateau
supernovae SN1999em. Although $P$ increased with time,  the
 polarization angle remained constant with time and wavelength indicating a common
 axis  of symmetry in the expanding envelope (Leonard et al. 2001, Wang et al. 2001).
}
\vskip -0.2cm
\label{obs}
\end{figure}

In recent years, there has been a mounting evidence that the explosions of massive stars (core
collapse supernovae) are highly aspherical. 
(1) The spectra of core-collapse supernovae (e.g., SN87A, SN1993J, SN1994I, SN1999em)
 are significantly polarized indicating asymmetric envelopes by factors up to 2
  (M\'endez et al. 1988; H\"oflich 1991; Jeffrey 1991; Wang et
al. 1996; Wang et al. 2001).
The degree of polarization  tends to vary inversely with the mass of the hydrogen
envelope, being maximum for Type Ib/c events with no hydrogen 
(Wang et al. 2000). For SN1999em (Fig. \ref{obs}),  Leonard et al. (2000) showed that the
polarization and, thus, the asphericity increases with time. Both trends suggest a connection
of the asymmetries with the central engine.
 For supernovae with a good time and wavelength
coverage, the orientation of the polarization vector tends to stay constant
both in time and in the wavelength.  This implies that there is a  global symmetry
axis in the ejecta (Leonard et al. 2001, Wang et al. 2001).
(2) Observations of SN~1987A showed that radioactive material was brought
to the hydrogen rich layers of the ejecta very quickly during the explosion
(Lucy 1988; Tueller et al. 1991).
(3) The remnant of the Cas~A supernova shows rapidly moving oxygen-rich
matter outside the nominal boundary of the remnant             
                 and  evidence for two oppositely directed jets of
high-velocity material (Fesen \& Gunderson 1997). 
(4). Recent X-ray observations with the CHANDRA satellite have shown an unusual
distribution of iron and silicon group elements with large
scale asymmetries in Cas~A (Huges et al. 2000).
(5) After the explosion, neutron stars are observed with high
velocities, up to 1000  km/s (Strom et al. 1995).
(6) Direct HST-images from June 11,2000, are able to resolve the inner debris of SN1987A showing
its prolate geometry with an axis ratio of $\approx 2$ (Fig. \ref{obs}).
Both the ejecta and the inner ring around SN1987A show a common axis of symmetry. By combining
the HST-images with spectral and early-time polarization data, Lifan Wang worked out the details of the
chemical and density structure. By connecting
the HST-images with the polarization data from earlier times, he demonstrated that
the overall geometry of the entire envelope of SN1987A was
 elongated all along the symmetry axis  (including the H-rich envelope), and that the distribution
of the products of stellar burning products (O, Ca, etc.) are concentrated in the equatorial plane
(Wang et al. 2001).

\section{Models for Collapse Supernova}
 
There is a general agreement that the explosion of a massive star is caused
by the collapse of its central parts into a neutron star or, for massive 
progenitors, into a black hole.
The mechanism of the energy deposition into the envelope
is still debated.  The process likely involves the bounce and 
the formation of the prompt shock (e.g. Van Riper 1978,
 Hillebrandt 1982), radiation of the energy in the form of
neutrinos (e.g. Bowers \& Wilson 1982), and the interaction of the neutrino with the material of
the envelope and  various types of convective motions ( e.g. Herant et al. 1994, Burrows et al. 1995,
 M\"uller \& Janka 1997, Janka \& M\"uller 1996), rotation (e.g. LeBlanc \& Wilson 1970, Saenz \& Shapiro S.L. 1981,
M\"onchmeyer et al. 1991, Zwerger \& M\"uller 1997)
and magnetic fields (e.g. LeBlanc \& Wilson 1970, Bisnovati-Kogan 1971).

 Currently, the most favored mechanism invokes the neutrino deposition.
 The results depend critically on the progenitor structure, equation of state,
 neutrino physics, and implementation of the neutrino transport.
Recent results indicate that spherical models fail to produce successful explosions
even when using sophisticated, relativistic Boltzmann solvers for the
neutrino transport and taking different flavors and neutrinos into account
(Yamada et al. 1999, Ramp \& Janka 2000, Mezzacappa et al. 2001).
 Multi-dimensional effects such as convection during the core collapse itself must 
be included but, still, it is an open question whether convection combined with the neutrino
transport provides the solution to the supernova problem (Ramp et al. 1998 and references therein).
In the current calculations, the size and scale of the convective motions 
seem to be too small to explain the observed asymmetries in the envelope. The angular
variability of the neutrino flux caused by the convection has been invoked to
explain the neutron star kicks (Burrows et al. 1995, Janka \& M\"uller 1994). Calculations give
kick velocity up to $\simeq 100$ km/s whereas NS with velocities of several
$100$ km/s are common.
 For a more detailed discussion of the current status and problems to explain 
the observed asymmetries,
see Khokhlov \& H\"oflich (2001).

It has long been suggested that the magnetic field can play an important role in the explosion
(LeBlanc \&  Wilson 1970; Ostriker \& Gunn 1971, Bisnovati-Kogan 1971, Symbalisty 1984).
LeBlanc and Wilson 
simulations showed the amplification of the magnetic field due to rotation and the
formation of two oppositely directed,
high-density, supersonic jets of material emanating from the collapsed
core. Their simulations assumed a rather high initial magnetic field $\sim 10^{11}$ Gauss
 and
produced a very strong final fields of the order of $\sim 10^{15}$ Gauss which seemed
to be unreasonable at the time. 
The recent discovery of pulsars with very high magnetic fields 
(Kouveliotou et al. 1998, Duncan \& Thomson 1992) revives the interest in
 the role of rotating magnetized neutron stars in the
explosion mechanism. It is not clear whether a  high initial magnetic field required 
for the LeBlanc \&  Wilson mechanism is realistic. On the other hand it may not be needed.

The current picture of the core collapse process is unsettled. 
A quantitative model of the core collapse must eventually include all the elements mentioned above.

Due to the difficulty of modeling core collapse from first principles, 
a very different line of attack on the explosion problem has been used extensively and 
proved to be successful
in understanding of the supernova problem, SN1987A in particular
 (Arnett et al. 1990, Hillebrand \& H\"oflich 1991).
The difference of characteristic time scales of the core (a second or less)
 and the envelope (hours to days) allows one to divide the explosion problem into two largely independent
parts - the core collapse and the ejection of the envelope. 
By assuming the characteristics of the energy deposition into the envelope 
during the core collapse, the response of the
envelope can be calculated. Thus, one can study the observational
consequences of the explosion and deduce characteristics of the core collapse and the progenitor structure.
This approach has been extensively applied in the framework of the 1D spherically symmetric
formulation. The major factors influencing the outcome have been found to be
the explosion energy and the progenitor structure. The same approach can be applied in multi-dimensions to
investigate the effects of asymmetric explosions.
 In this paper we study the effects and observational consequences of an asymmetric, jet-like
deposition of energy inside the envelope of SN.

\section{Results for Jet-Induced Supernovae}
\vskip -0.7cm
 
\noindent
\subsection{Numerical Methods}
 
{\bf 3-D Hydrodynamics:}  The explosion and jet propagation are
 calculated by a full  3-D code within a cubic domain.
The stellar material is  described by the time-dependent, compressible,
Euler equations for inviscid flow with an ideal gas equation 
with $\gamma=5/3$ plus a component due to radiation pressure with 
$\gamma=4/3$.  The Euler
equations are  integrated using an explicit, second-order accurate, Godunov type,
adaptive-mesh-refinement, massively parallel, Fully-Threaded Tree (FTT)
program, ALLA  (Khokhlov 1998).

{\bf 1-D  Radiation-Hydrodynamics:}                                           
 About 1000 seconds after the core collapse and in case of the explosion of red supergiants,
the propagation of the shock front becomes almost spherical (see below). To be able to follow
the development up to the phase of homologous expansion ($\approx 3-5$ days), the 3-D structure
is remapped on a 1-D grid, and the further evolution is
 calculated using a one-dimensional radiation-hydro code  that solves the hydrodynamical
equations explicitly in the comoving frame by the piecewise parabolic method 
(Colella and Woodward 1984). The radiation transport part is solved implicitly using the method
of variable Eddington factors.  Expansion opacities and LTE-equations of state are used
 (H\"oflich et al. 1998 and references therein).

{\bf 3-D Radiation Transport for  Ionization Structures and Light Curves:}
 For given, arbitrary 3-D  density, velocity and chemical distributions,
we calculated the detailed 3-D ionization structure and LCs using the same assumptions as for our
1-D hydro code. The $\gamma$-ray transport is computed in 3-D using
a Monte Carlo method (H\"oflich et al. 1992, 1993) which includes
relativistic effects and a consistent treatment of both the
continua and line opacities. Subsequently, for low energy photons, we solve the
three-dimensional radiation transport using a
hybrid scheme of Monte-Carlo and non-equilibrium diffusion methods. 
 As a first step, we solve the time-dependent radiation transport equation
in non-equilibrium, diffusion approximation for 3-D geometry including the  
scattering and thermalization terms for the source functions, and include the 
frequency derivatives into the formulation for the opacities and emissivities.
 We solve the same set of momentum equations as H\"oflich et al. (1993)
 but for 3-D geometry and an
Eddington factor of 1/3. At large optical depths, this provides the solution for the
full radiation transport problem. In a second step,
 to obtain the correction  solution for the radiation transport equation at small optical depths, the difference
between the solution of the diffusion and full radiation transport equation is calculated
in a  Monte Carlo scheme. We calculate the difference between the solutions for computational
accuracy and efficiency.
 Consistency between the solution at the outer and inner region is obtained  iteratively.
The same Monte Carlo solver is used which has been applied to compute $\gamma$-ray
and  polarization spectra. For simplification, the relation between energy 
density and temperature is taken from the 1-D radiation hydrodynamics at the corresponding 
time (see below). 

{\bf 3-D Radiation for Spectra and Polarization:}
 For several moments of time, detailed polarization and  flux spectra for asymmetric explosions
are calculated using our MC-code including detailed equations of state, and detailed atomic models for
some of the ions. For details, see H\"oflich (1995), H\"oflich et al. (1995) and Wang et al. (1998).

\begin{figure}[th]
\includegraphics[width=8.9cm,clip=,angle=0]{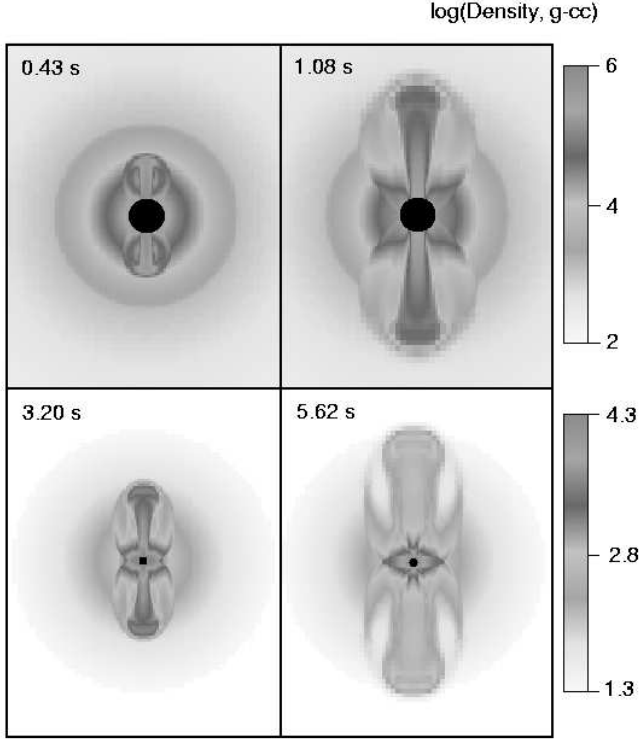}
\caption {Logarithm of the density structure 
as a function of time for a helium core.
The total mass of the ejecta is 2.6 $M_\odot$.
The initial radius, velocity  and density of the jet were taken to 1200 km
32,000 km/sec and $6.5E5 g/cm^3$, respectively. The shown domains 7.9,       
9.0, 36 and 45 $\times 10^9 cm $.
The total energy is about 9E50 erg. After about 4.5 seconds, the 
jet penetrates the star. The energy deposited in the stellar envelope by the jet
is about 4E50 erg, and the final asymmetry is of the order of two.}
\vskip -0.2cm
\label{jet1}
\end{figure}
 
\begin{figure}[b]
\includegraphics[width=4.0cm,angle=270]{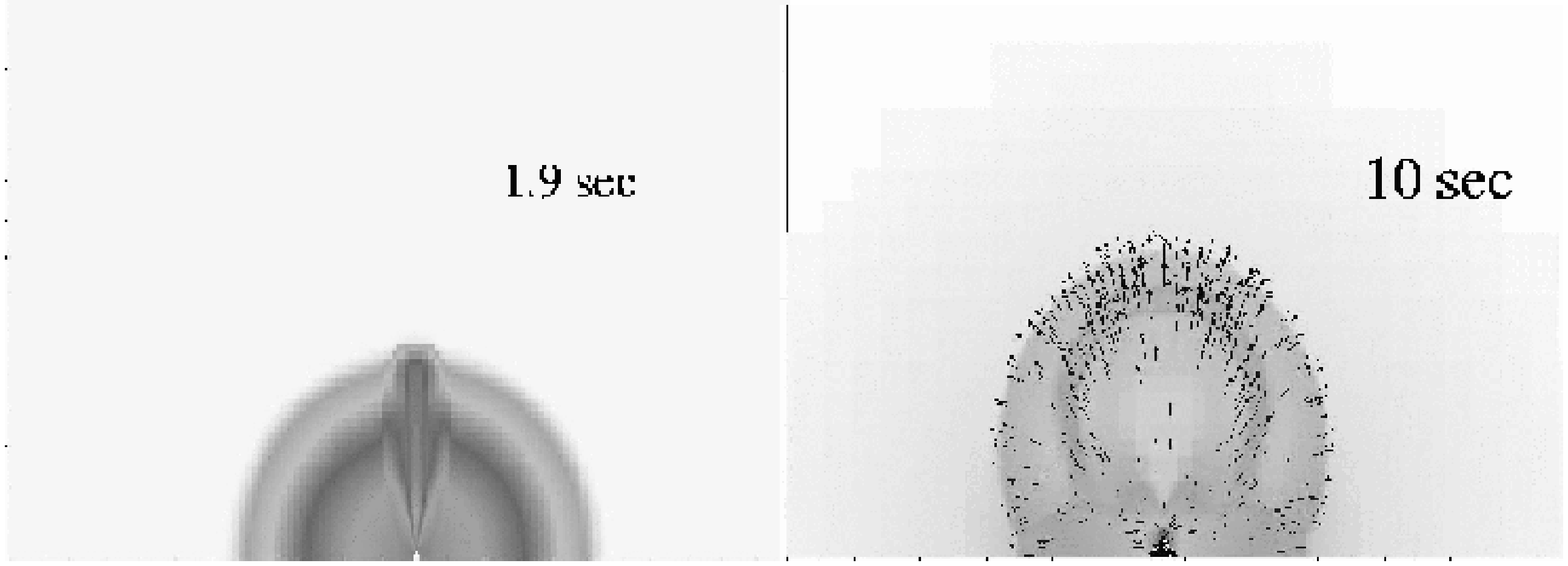}
\caption {Same as Fig. \ref{jet1} ($0.5 \leq log(\rho)\leq 5.7 $)
 but for a jet velocity of 61,000 km/sec and
a total energy of 10 foe at $\approx 1.9 sec$ (left), and 
 11,000 km/sec and a total energy of 0.6 foe (right).  The size of the
presented domains are 5 (left) and 2 $10^{10} cm$ (right), respectively.
 For the high velocity jet,
 most of the energy is carried away by the
jet. Only 0.9 foe are deposited in the expanding envelope.
 In case of a low velocity jet, the bow-shock still propagates through
the star after the jet is switched off,                   and
the entire jet energy is deposited in the expanding envelope.}
\vskip -0.2cm
\label{jet2}
\end{figure}
 
\subsection { Jet propagation:}
 
 {\bf The Setup:}
   The  computational domain is a cube of size $L$    
with a spherical star of radius $R_{\rm star}$ and mass $M_{\rm star}$
 placed in the center.
 The innermost part with mass $M_{\rm core} \simeq 1.6 
M_{\odot}$ and radius  $R_{\rm core} = 4.5    \times  10^8$~cm, consisting
of Fe and Si, is assumed to have collapsed on a timescale much faster
than the outer, lower-density material. It is removed and replaced by a
point gravitational source with mass $ M_{\rm core}$ representing the
newly formed neutron star.  The remaining mass of the envelope $M_{env}$
is mapped onto the
computational  domain. 
 At two polar locations where the jets
are initiated at $R_{core}$, we impose an inflow with velocity $v_j$
$\rho_j$.   
 At  $R_{\rm core}$,
the jet density and pressure are the same as those of the background
material.
For the first 0.5~s, the jet velocity at $R_{\rm core}$ is kept
constant at $v_j$.
  After 0.5~s, the
velocity of the jets at $R_{\rm core}$ was gradually decreased to zero
at approximately 2~s. The total energy of the jets is $E_j$.
 These parameters are
consistent within, but somewhat less than, those of the LeBlanc-Wilson model.

{\bf The reference model:} As a baseline case, we consider a jet-induced explosion in a helium
star. 
  Jet propagation inside the star is shown in
Fig. \ref{jet1}.
 As the jets move outwards, they remain collimated and do not  develop much internal
structure. A bow shock forms at the head of the jet and spreads in all
directions, roughly cylindrically around each jet.   
  The jet-engine has been switched off after about 2.5 seconds
the material of the bow shock continues to propagate through the star.                    
 The stellar material is shocked by the bow shock. Mach shocks
travels two wards the equator resulting in a redistribution of the 
energy. The opening angle of the jet depends on the ratio between the
velocity of the bow shock to the speed of sound. For a given star,    
 this angle determines the efficiency of the deposition of the
 jet energy into the stellar envelope. Here, the
efficiency of the energy deposition is about 40 \%, and
the final asymmetry of the envelope is about two.

\noindent
 {\bf Influence of the  jet properties:}
 Fig. \ref{jet2} shows two examples of an explosion with
with a low and a very high jet velocity compared to the baseline case (Fig. \ref{jet1}).
  Fig. \ref{jet2}  demonstrates the influence of the jet velocity on the opening
 angle of the jet and, consequently,
 on the efficiency of the energy deposition. For the low velocity jet,
the jet engine is switched off long before
the jet penetrates the stellar envelope. Almost all of the energy of the
jet goes into the stellar explosion. On a contrary, the  fast jet (61,000 km/sec)
 triggers only a weak explosion of 0.9 foe although its 
 total energy was $\approx 10 foe $.

{\bf Influence of the progenitor:}
 For a very extended star, as  in  case of 'normal'  Type II Supernovae,
the bow shock of a low velocity jet stalls within the envelope, and
 the entire jet energy is used to trigger the ejection of the stellar envelope. In
our example (Fig. \ref{model}),
the jet material penetrates the helium core at about 100 seconds.
 After about 250 seconds the material of the jet  stalls within the
hydrogen rich envelope and
 after passing about 5 solar masses in the radial
mass scale of the spherical progenitor.
 At this time, the isobars are almost spherical,  and an almost 
spherical shock front travels outwards. Consequently, 
 Strong asphericities are 
limited to the inner regions.
 After about 385 seconds,  we stopped the 3-D run and remaped the  outer layers into 
1-D structure, and followed the further evolution in 1-D.
 After about 1.8E4 seconds, the shock front reaches 
the surface. After about 3 days, the envelope expands homologously.
The region where the jet material stalled, expands at velocities of about
4500 km/sec.

{\bf Fallback:}
 Jet-induced supernovae have very different characteristics with 
respect to fallback of material and the innermost structure.
 In 1-D calculations and for stars with
Main Sequence Masses of less than  20 $M_\odot$ and explosion
energies in excess of 1 foe, the fallback of material remains less than
1.E-2 to 1.E-3 $M_\odot$ and an inner, low density cavity is formed 
with an outer edge of $^{56}Ni$. For explosion energies between 1 and 2 foe,
 the outer edge of the cavity expands typically with
velocities of about 700 to 1500 km/sec
(e.g. Woosley 1997,  H\"oflich et al. 2000).
 In contrast, we find strong, continuous fallback of     
$\approx 0.2  M_\odot$ in the  the 3-D hydro models,
and no lower limit for the velocity of  the expanding material (Fig. 4 of Khokhlov \& H\"oflich 2001).
 This significant amount of fallback  must have important consequences for the
secondary formation of a black hole.
 The exact amount and time scales for the 
final accretion on the neutron star will depend sensitively on the
rotation and momentum transport.

{\bf Chemical Structure:}
 The final chemical profiles of elements 
formed during the stellar evolution  such as He, C, O and Si
are 'butterfly- shaped' whereas the jet material fills
an inner, conic structure (Fig. \ref{model}, upper, middle panel).
 
 The
composition of the jets must reflect the composition of the innermost
 parts of the star, and should contain heavy 
and intermediate-mass elements, freshly synthesized material such
as $^{56}Ni$ and, maybe, r-process elements because, in our examples,
the entropy at the bow shock region of the jet was as high as a few hundred.
In any case,
during the explosion, the jets bring heavy and intermediate mass elements
into the outer H-rich  layers.

\subsection{Radiation Transport Effects}
  
\begin{figure}[ht]
\includegraphics[width=4.0cm,angle=270,clip=]{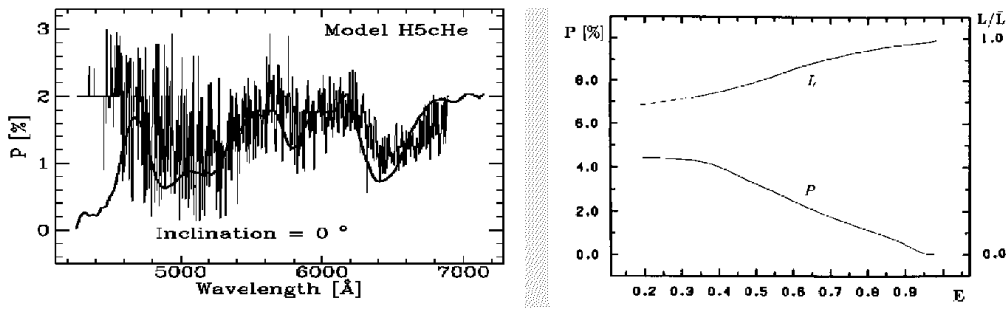}
\caption {
Polarization spectrum for SN1993J for an axis ratio of 1/2 for an oblate ellipsoid
in comparison with observations by
Trammell et al. (1993) are given in the left plot. On the right,
the dependence of the continuum polarization (right) and directional
dependence of the luminosity is shown 
 as a function
axis ratios for oblate  ellipsoids  seen from the equator
(from H\"oflich, 1991 \& H\"oflich et al. 1995).
}
\vskip -0.2cm
\label{pol2}
\end{figure}

 For the compact progenitors of SNe~Ib/c, the final departures of the 
iso-density contours from sphericity  are typical a factor of two. This 
 will produce a linear polarization of about
2 to 3 \% (Fig. \ref{pol2}) consistent with the values observed for Type Ib/c supernovae.
 In case of a red supergiant, i.e. SNe~II,
the   asphericity is restricted to the inner layers of the H-rich envelope. There
the iso-densities show an axis ratios  of up to $\approx$ 1.3.
 The intermediate and outer H-rich layers remain spherical.
 This has strong  consequences for the observations,  in particular,
for polarization measurements. In general, the polarization should be
larger in SNe~Ib/c compared to classical SNe~II which is consistent
with the observations by Wang et al. (2000).                    
    Early on, we expect no or little
polarization in supernovae with a massive, hydrogen rich envelope which
will increase with time to about 1\%                            
(H\"oflich 1991), depending on the inclination the SN is 
observed. This is also consistent both with the long-term time evolution
of SN1987A (Jefferies 1991).

 Recently, the plateau supernova 1999em has been observed
with VLT and Keck  providing the
best time coverage up to know of any supernovae
(Wang et al. 2000; Leonard et al. 2001). The basic trend has been confirmed which we expected from
the hydro. Indeed,
$P$ is very low early on, and it rises when more central parts are seen.
 However, there are profound differences which point towards an additional mechanism
to produce aspherical photospheres.
\begin{figure}[h]     
\includegraphics[width=7.5cm,angle=270,clip=]{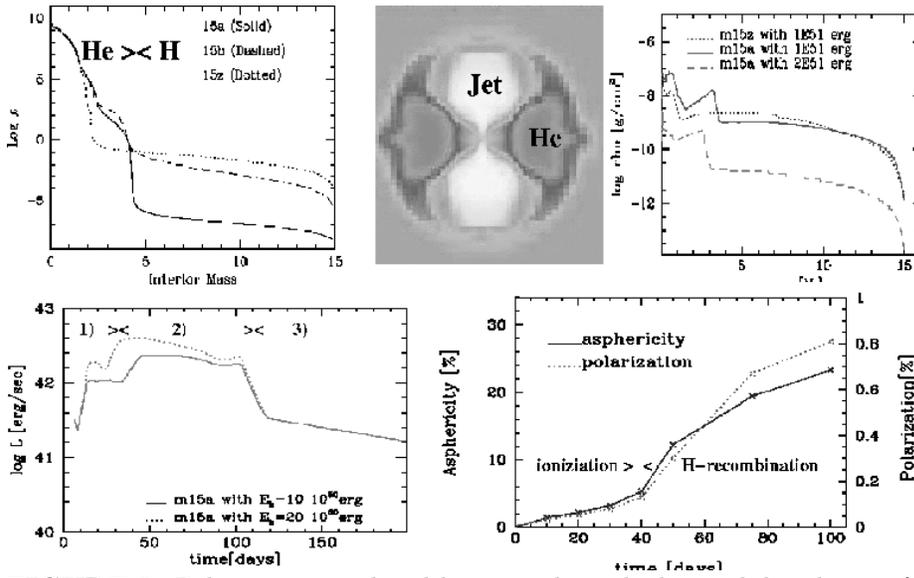}
\caption {Polarization produced by an aspherical, chemical distribution
for a SN~IIp model with $15 M_\odot$ and an explosion energy $E_{exp} = 2 \times 10^{51}erg$.
 This model resembles  SN~1999em (see above).
The initial density profile is given for a star at the final stage of stellar evolution
for metallicities Z of 0.02, 0.001 and 0 (models 15a, 15b, 15z, upper left panel,
from H\"oflich et al. 2000 \& Chieffi et al. 2001). The model for the  Red Supergiant,
 15a, has been used to calculation of the jet-induced explosion in sect. 2).
 In the upper, middle panel, the chemical distribution of He
  is given at 250 sec
for the He-rich layers after the jet material has stalled.
The color-codes white, yellow, green, blue and red correspond to He mass fractions of
0., 0.18, 0.36, 0.72, and  1., respectively.
 The subsequent explosion has been followed in 1-D up to the phase of homologous
expansion. In the upper, right panel, the density distribution is given at about 5 days
after the explosion.  The steep gradients in the density in the upper right and left panels
are located at the interface between the He-core and the H-mantel.
 In the lower, left panel,
the resulting bolometric LCs are given for a our $E_{exp}=$
 $2 \times 10^{51}erg$ (dotted line) and, for comparison, for $1\times 10^{51}erg$, respectively.
 Based on full 3-D calculations for the radiation \& $\gamma $-ray transport,
we have calculated the location of the recombination front as a function of
time. The resulting shape of the photosphere is always prolate.
 The corresponding axis ratio and the  polarization seen from the equator are shown 
 (lower, right panel).
Note the strong increase of the asphericity after the onset of the recombination phase between
day 30 to 40  (see also SN~1999em in Fig. \ref{obs}).
}
\vskip -0.2cm
\label{model}
\end{figure}

  SN1999em  is an extreme plateau supernovae with a plateau lasting for more than 100 days
(IAUC 7294 to 7355). However, no detailed light curves have been published.  Therefore,
in Fig. \ref{model}, we show a theoretical LC which resembles SN~1999em with respect to the
duration of the plateau and its brightness, and the typical expansion velocities.
The light curves of SNe~IIp show three distinct phases 
 (Fig. \ref{model}).
1)  Most of the envelope is ionized. This phase
depends sensitively on the explosion energy, mixing of radioactive Ni, and the mass of the progenitor,
 e.g. either strong mixing or $E_{kin}\leq 1 foe $ will cause a steep and steady increase in the
luminosity (and in B and V);
 2) The emitted energy is determined by the receding (in mass)
  of the H recombination front which is responsible for both the release of stored, thermal and
the recombination energy. At the recombination front, the opacity drops by about 3 orders
of magnitude when it changes from electron scattering dominated to bound-free/ free-free.
This provides a self regulating mechanism for the energy release. If too little energy is released, the opacity
drops fast causing an increase in the speed at which the  the photosphere is receding.
 In term, this causes a larger energy release and vs. . Hydrogen recombines at a
specific temperature at or just below the photosphere.
Due to the flat density profiles of the expanding envelopes in the RSG case,
 the photospheric radius and, thus, the luminosity L stays almost constant.
  After the recombination front has
passed through the H-rich envelope,  the brightness drops fast.
During phase 3), L is given by the instant energy release by radioactive decay of $^{56} Co$.
 Obviously, the steep rise in $P$ of SN1999em coincide with the transition from phase 1 to 2,
pointing towards un-isotropic excitation as a new mechanism for producing aspherical photospheres and,
consequently, polarization. To quantify the effect, we have calculated the temperature
and ionization structure of a SN~IIp (Fig. \ref{model}). Starting from a spherical model,
the initial chemical distribution has been taken from our 3-D jet simulation. As mentioned above,
the chemical profile is frozen out after about 250 sec, and the expansion becomes spherical.
 The further evolution can be followed with our 1-D radiation code. For several moments of time,
we have calculated the ionization structure and continuum polarization based of the 3-D chemical
structure and the spherical density distribution under the assumption that the distribution of the
radioactive Ni coincides with the jet-material. Note that the $^{56}Ni$ layers extend throughout
the He-core ($\approx 5 M_\odot$) in polar directions but they are confined to the very center
along the equatorial plane (Fig. \ref{model}, upper, middle panel) leading to an increased transport
of energy  and, consequently, heating toward polar directions. Before the recombination phase,
the opacity is dominated by Thomson scattering which does not depend on the temperature, and the
shape of the photosphere remains almost spherical. However, during the recombination phase,
the location of the photosphere depends sensitively on the heating, and the photosphere becomes
prolate (Fig. \ref{model}, lower, right panel). There is a gradual increase in P and no
jump $P$  because of the optically thick H-rich layers below the photosphere redistribute the
photons.  The increase of P depends on the geometrical expansion and speed (in mass) of the receding
recombination front.

\section{Conclusions}

We have numerically studied the explosion of Core Collapse 
supernovae caused by supersonic jets generated in the center of the 
supernova as  a result of the core collapse into a neutron star. 
We  simulated  the process of the jet propagation through the star, 
the redistribution of elements, and radiation transport effects.
 A  strong explosion and a high efficiency for the conversion of the
jet energy requires low jet velocities or a low, initial collimation
of the jet.  With increasing extension of the envelope, the 
conversion factor increases. Typically, we would expect higher
kinetic energies in SNe~II compared to SNe~Ib/c if a significant amount
of explosion energy is carried away by  jets. 
Within the framework of jet-induced SN, the lack of this evidence
suggests that the jets have low velocities.
 
 The He, C, O and Si rich layers                     of the 
progenitor show   characteristic, butterfly-shape structures.
This  overall morphology and pattern should be observable in supernovae remnants, e.g. with the Chandra 
observatory despite some modifications and instabilities when the expanding medium interacts with the 
interstellar material.
 
   During the explosion, the jets bring heavy and intermediate mass elements
into the outer layers including $^{56}Ni$. Due to the high entropies
of the jet material close to the center, this may be a possible site
for r-process elements.
Spatial distribution of the jet material will influence the
properties of a supernova.
 In our model for a SN~II, the jet material stalled within the expanding envelope 
corresponding to a velocity of $\approx 4500 km/sec$ during
the phase of homologous expansion.
 In SN1987A, a bump in spectral lines of various elements has been
interpreted by material excited by a clump of radioactive $^{56}Ni$
(Lucy 1988). 
Within our framework, this bump may be a measure of 
region where the jet stalled.  
This could also explain the early appearance of X-rays
 in SN1987A which requires strong mixing of radioactive material
into the hydrogen-rich layers
  (see above), and the overall distribution of elements and distribution of elements
in the resolved HST-images of the inner debris of SN~1987A.
   We note that, if this interpretation
is correct, the 'mystery spot' (Nisenson et al. 1988) would
be unrelated.
 In contrast to 1-D simulations, we find in our models strong,  
continuous fallback over an extended period of time, and a lack 
of an inner, almost empty cavity. This significant amount of fallback
and the consequences for the secondary formation of a black hole shall
be noted. Moreover, fallback and the low velocity material may alter
the  escape probability for $\gamma $-rays produced by radioactive 
decay of $^{56}Ni$. In general, the lower escape probability is           
unimportant for the determination of the total $^{56}Ni$ production by the
late LCs   because full thermalization can be assumed in core collapse
SN during the first few years.
 However, in extreme cases such as SN98BW (e.g.
Schaefer et al. 2000), only
a small fraction of gamma's are trapped. Effects of multi-dimensionality 
will strongly alter the  energy input by radioactive material and
disallow a reliable estimate for the total $^{56}Ni$ mass.

 Qualitatively, the jet-induced picture allows to reproduce the 
polarization observed in core collapse supernovae.
 Both asymmetric ionization and density/chemical distributions have been identified
as crucial. Even within the picture of jet-induced explosion, the latter effect alone 
cannot (!) account for the high polarization produced in the intermediate H-rich layers
of core-collapse SN with a massive envelope such as  SN~1999em.
 The former effect operates only during a recombination phase, and can be expected
to dominate the polarization in core-collapse supernovae with massive H-rich envelope during 
the first 1 to 2 months. Complete time series of polarization measurements are needed to test
this suggestion.

 Finally, we want to emphasize the limits of this study and some of the
open questions which will be addressed in future.
 We have assumed that jets are formed in the course 
of the formation of a neutron star, and have addressed observational
 consequences and constrains. However, we have not calculated the jet
formation, we do not know if they really form, and, if they form, whether
they form in all core-collapse supernovae.
Qualitatively, the observational properties of
core collapse supernovae are consistent with jet-induced supernovae
and support strongly that the explosion mechanism is highly aspherical but more
detailed comparisons with an individual objects must be performed as soon as the 
data become available. 
We cannot claim that the jets are the only mechanism that can explain
asphericity in supernovae but any competing mechanism must involve some sort of
axial symmetry on large scales with a profound impact on the explosion such as 
rapid rotation.  It remains to be seen whether asymmetry and  axial symmetry are 
the 'smoking gun' for our understanding of the SN-mechanism.                                           mm

\noindent{\sl Acknowledgments:} We want to thank our colleagues for helpful discussions,
in particular, D. Baade, E.S. Oran, J.C. Wheeler,  Inzu Yi A.,
C. Mayers, J.C. Wilson, A. Chieffi,
 M. Limongi, and O. Straniero. This work is supported in part by NASA Grant                  
LSTA-98-022.

\end{document}